# A Statistical Estimation of the Occurrence of Extraterrestrial Intelligence in the Milky Way Galaxy

Xiang Cai[1], Jonathan H. Jiang[2], Kristen A. Fahy[2], Yuk L. Yung[3]

1. Santiago High School, Corona, CA
2. Jet Propulsion Laboratory, California Institute of Technology, Pasadena, CA
3. Division of Geological and Planetary Sciences, California Institute of Technology, Pasadena, CA

## Abstract

In the field of Astrobiology, the precise location, prevalence and age of potential extraterrestrial intelligence (ETI) have not been explicitly explored. Here, we address these inquiries using an empirical galactic simulation model to analyze the spatial-temporal variations and the prevalence of potential ETI within the Galaxy. This model estimates the occurrence of ETI, providing guidance on where to look for intelligent life in the Search for ETI (SETI) with a set of criteria, including well-established astrophysical properties of the Milky Way. Further, typically overlooked factors such as the process of abiogenesis, different evolutionary timescales and potential self-annihilation are incorporated to explore the growth propensity of ETI. We examine three major parameters: 1) the likelihood rate of abiogenesis ($\lambda_A$); 2) evolutionary timescales ($T_{evo}$); and 3) probability of self-annihilation of complex life ($P_{ann}$). We found $P_{ann}$ to be the most influential parameter determining the quantity and age of galactic intelligent life. Our model simulation also identified a peak location for ETI at an annular region approximately 4 kpc from the Galactic center around 8 billion years (Gyrs), with complex life decreasing temporally and spatially from the peak point, asserting a high likelihood of intelligent life in the galactic inner disk. The simulated age distributions also suggest that most of the intelligent life in our galaxy are young, thus making observation or detection difficult.

1. **Introduction**

For centuries, there have been fundamental questions regarding intelligent life within the Galaxy. The study of extraterrestrial life intelligence (ETI) has long been based on only one sample: Earth. The Search for ETI (SETI, e.g., Tarter, 2001) has been heavily limited by our current technology and a lack of quantitative understanding of intelligent life on a galactic scale, and the answers to the absence of life discovery elsewhere have remained controversial. For instance, Hart (1975) suggested that no other intelligent life exists to address the question of the Fermi Paradox, which contradicts a lack of evidence and high estimates of ETI. However, many scholars challenge this stance, as it was based on the argument that SETI is extensive enough to exclude ETI elsewhere (Carroll-Nellenback *et al.*, 2019); this is invalid, as we cannot *a priori* preclude the possibility of the existence of other intelligent life in the Galaxy. Therefore, it is warranted to develop a spatial-temporal analysis on the quantitative propensity of ETI, and a galactic-scale evaluation of different key parameters to explore the likely characteristics of intelligent life within the Milky Way. Moreover, while relying heavily on Earth as one sample may not seem ideal, we are not devoid of other information; with the recent investigation of stellar observations, we can apply our knowledge from Earth to Earth-like planets to develop a comprehensive overview of where and when intelligent life may emerge and develop in the Galaxy.

The quantitative approaches taken to address this challenge have been largely influenced by the application of the Drake equation (Drake, 1965); a simple algebraic expression to investigate the quantity of intelligent life and its simplistic nature exposes it to many re-expressions (Walters *et al.*, 1980; Burchell, 2006; Forgan, 2009). Despite its benefits, the equation lacks a temporal dependence (Cirkovic, 2004), and does not consider evolving stellar properties in the Milky Way. Additionally, there are inherent parameter uncertainties (Maccone, 2010; Glade *et al.*, 2012), resulting in highly subjective estimations.

The existence of extraterrestrial intelligence is directly related to habitability and a galactic habitable zone (GHZ); where habitable planets are located and where potential life are most likely to form. Gonzalez *et al.* (2001) first proposed spatial and temporal aspects of galactic habitability and assessed it by quantifying the abundance of metals in the Galaxy and predicting favorability for planet formation. This study was expanded by Lineweaver *et al.* (2004), who discussed catastrophic transient radiation events, the formation of a Hot Jupiter, and estimated the GHZ to be an annular region of galactocentric radius between 7-9 kpc that widens with time. Further, with the same prerequisites for complex life as Lineweaver *et al.* (2004), Gowanlock *et al.* (2011)

developed a Monte Carlo simulation of habitable planets that sampled individual stars to investigate where habitable stars are located, adopting the galactic stellar attributes of an inside-out star formation history (Naab and Ostriker, 2006) and a three-dimensional stellar number density (Jurić *et al.*, 2008). Based on this model, they predicted that most habitable planets are likely to be in the inner Galaxy, with a higher fraction of stars hosting a habitable planet. However, Prantzos (2008) suggested an alternative perspective that the entire Galaxy's disk may be habitable, as the necessary conditions rendering the inner Galaxy habitable have not been confirmed.

Further, inspired by the former research on habitability, inquiries upon intelligent life within the Milky Way have been developed with Monte Carlo techniques. Extending the Monte Carlo modeling from Gowanlock *et al.* (2011), Morrison and Gowanlock (2015) explored the opportunities of ETI by considering the extended timescale required for life to evolve into intelligence. They posited that the gaps between the transient radiation events that allow for evolution can vary greatly based on a number of spatial-temporal factors. Supporting results from Gowanlock *et al.* (2011), they concluded that the maximum number of opportunities occur in the inner Galaxy despite a higher frequency of sterilizing events. However, they were unable to provide a precise estimation for the peak location within their limited galactic range. Similarly, Forgan (2009) and Forgan and Rice (2010) described a Monte Carlo model estimating the likelihood of life on a habitable planet will evolve into intelligence by adopting a simplified model that resets life in regular intervals. Hairs (2011) utilized a Gaussian distribution to model the prevalence of intelligence and the inter-arrival times of successive intelligence; however, it too lacked spatial and temporal analysis.

Overall, previous studies of habitability and the likelihood of intelligent life have provided many valuable insights to the practice of SETI; however, given the many uncertainties involved, the precise propensity of galactic intelligent life to emerge has not yet been explored with spatial and temporal analysis, nor has any research explicitly estimated an age distribution for potential life within the Galaxy. Furthermore, three major components have remained largely untouched: 1) the probability of life in a pre-biotic condition, 2) different potential timescales for biological evolution, and 3) the probability of self-annihilation of complex life. The work presented herein will incorporate these important, yet often neglected components, in our galactic model.

To approach our questions regarding the emergence of life, the most direct and promising way is to search for life on other planets in and beyond the solar system. With recent stellar observations and computational power, a theoretical approach is possible. Inspired by earlier work (Spiegel and Turner, 2012; Scharfa and Cronin, 2016; Chen and Kipping, 2018; Kipping, 2020), we treated the process of abiogenesis as a Poisson process to acquire the probability of life arising in a pre-biotic condition as a product of time and a uniform rate ($\lambda$). Using our model, if we evaluate how much the quantity of complex life changes in the Milky Way within a specific time range, we could have a better understanding on the hypothesis of life origins, the *Rare Earth Hypothesis*.

The timescale for biological evolution has been largely assumed to be identical to the intellectual species on Earth (e.g., Lineweaver *et al.*, 2004; Gowanlock *et al.*, 2011). Without considering the possibility that life may require varied timescales to evolve (component 2), any result of the growth propensity of life is dependent on the ongoing debate whether or not humans on Earth are the paradigm resembling all other evolving complex life. Many scholars caution against considering this a typical timescale for evolution (e.g., Livio, 1999; Carter, 2008). To assess this uncertainty, we selected different potential timescales and present the variant results to explore the influence of this parameter and all possible outcomes.

Additionally, the potential possibilities of self-annihilation during the development of complex life (component 3) has been heavily disregarded (e.g., Morrison and Gowanlock, 2015). Though some studies have attempted to include this perspective (Forgan, 2009; Forgan and Rice, 2010), there has been no galactic examination on the influence of this parameter on the quantity of ETI with spatial-temporal considerations; thus, the subjective nature of this subject remains untouched.

The questions of ETI should be approached in a more quantitative and scientific way, taking considerations from both astrophysical and biological perspectives to explore the emergence, development, prevalence, and age distribution of complex life in the Milky Way over time and location, and provide a complement to SETI with direct lines of investigation. When exploring the spatial-temporal prevalence of intelligent life with the parameters presented above, it not only allows for the reduction of speculations, but also gauges the possible resolutions of the Fermi Paradox.

## 2. Methodology

We developed a Lagrangian model with Monte Carlo techniques to accurately simulate each Earth-like planet in the Milky Way. This model incorporated a Rust ECS (Entity-Component-System) Specs Framework to effectively manage each star and planet particle, and spatial hashing technique to optimize the large-scale, real-time simulation. Our simulation modeling approach to the ETI investigation is illustrated in the Appendix Figure 1.

The overall simulation method is summarized as follows:

1. Initiated a 3D spatial hash table of Milky Way with distributed gas mass;
2. Generated Sun-like stars harboring Earth-like planets and activated supernova explosion with the same distribution as observations;
3. For each Earth-like planet, allow life to emerge with the Poisson process of abiogenesis;
4. For each life-bearing planet free from transient events (e.g., supernova), follow life's evolution into intelligence.

**2.1 Formation of Sun-like Star Harboring Earth-like Planet**

*2.1.1 Prevalence of Sun-like Star Harboring Earth-like Planets*

Within our simulation, we defined Sun-like stars as stars with mass of 0.8 $M_\odot \leq M \leq 1.2$ $M_\odot$, where $M_\odot = 1$ denotes the mass of the Sun. We also adopted the initial mass function (IMF, the initial distribution of mass for a star population), with $\alpha = 2.35$ from Salpeter (1955), and calculated the approximate Sun-like star fraction in the Milky Way. Next, we described the Earth-like planets as planets: (1) with 1-2 Earth-radii, (2) receiving stellar energy within a factor of 4 compared to that of Earth, and (3) with Earth-like orbital period of 200 to 400 days. From these perspectives, we used recent observational analysis on the prevalence of Earth-sized planets from Petigura *et al.* (2013), who estimated that 11% of the Earth-size planets around Sun-like stars receive Earth-like stellar energy, and 5.7 % of the Earth-sized planets obtain Earth-like orbital periods. We multiplied these probabilities to acquire the fraction of Sun-like stars harboring Earth-like planets in the Galaxy that fit our four criteria described above.

*2.1.2 Spatial Hash Table and Distribution of Gas*

We designed a galactic spatial hash table with the same volume of the Milky Way, uniformly divided into three-dimensional cells with the volume of each cell at 0.01 $kpc^3$. This galactic table enables the computation of large-scale objects at a real-time frame rate, and allows fast location and proximity detection queries (Hastings *et al.*, 2005). Recent work by Fuentes *et al.* (2017) used a similar approach to implement the hierarchical density in investigating galactic formation and evolution.

To simulate the formation of stars (the collapse of cold gas under its own weight in interstellar space), we calculated the mass of gas for each cell, and the initial distribution of gas at the Galaxy's formation (1). The gas surface density decreases exponentially with the galactocentric radius (Naab and Ostriker, 2006):

$$\Sigma_{gas}(r, 0) = \Sigma_c e^{-r/h_R} \qquad (1)$$

where the quantity $\Sigma_{gas}$ is the surface density of gas ($M_\odot$ kpc$^{-2}$), and r is the radial distance from the Galactic Center (kpc). The constant $\Sigma_c$ $1.4 \times 10^8$ $M_\odot$ pc$^{-2}$ is the central surface density of the Galaxy (Donato et al., 2009), and $h_R$ = 2.25 kpc denotes the radial disk scale length. At the time of the simulated Galaxy's formation, this equation assigns different total surface densities of gas to each cell based on their distance from the Galactic Center.

*2.1.3 Upper Limit of Star Formation*

To enable star formation from interstellar gas, we utilized the Schmidt-Kennicutt law (Schmidt, 1959), which is a simple relationship between gas surface density and star formation rate (SFR) surface density:

$$\Sigma_{SFR} = A \Sigma_{gas}{}^N \qquad (2)$$

where the quantity of $\Sigma_{SFR}$ represents the surface density of SFR ($M_\odot$ kpc$^{-2}$ Myr$^{-1}$), and $\Sigma_{gas}$ is the dimensionless $\Sigma_{gas}$ (surface density of interstellar gas) in the unit of $M_\odot$ kpc$^{-2}$. The constant A = 250 $M_\odot$ kpc$^{-2}$ Myr$^{-1}$ (Kennicutt, 1998), and N = 1.4 (Kennicutt, 1998; Arifyanto *et al.*, 2005). For each cell within the galactic model, we first calculated its $\Sigma_{SFR}$ from the initially-assigned $\Sigma_{gas}$ described above, and then calculated the theoretical stellar mass of SFR ($M_{SFR}$, in $M_\odot$ Myr$^{-1}$) by multiplying its surface area. We estimated the $M_{SFR}$ for Sun-like stars harboring Earth-like planets according to the fraction from §*2.2.1*. For each time step (Myr), our galactic model accumulates the theoretical stellar mass available for the formation of stars ($M_{SF}$, in $M_\odot$). This means, at each location, newly available stellar mass from the collapse of gas will be added to the total available $M_{SF}$ according to the $M_{SFR}$. We then acquired the theoretical stellar mass for the star formation and used it as an upper limit for the total stellar mass of star formation in each cell within the galactic model. This theoretical estimate is not intended for the SFR, but rather, to incorporate the relationship between gas and star formation. We utilized the SFR according to a recent chemical evolution model of star formation history (SFH), and the process of star formation will be described in detail in §*2.1.5*.

*2.1.4 Stellar Mass and Main Sequence Lifetime*

For each star, we uniformly assigned the mass ranges from 0.8 $M_\odot$ ≤ M ≤ 1.2 $M_\odot$ and its main sequence lifetime, the star's overall lifespan, corresponding to its mass (Hansen and Kawaler, 1994):

$$T_L = T_{L\odot} (M_\odot /M)^{2.5} \tag{3}$$

where $T_{L\odot} = 11,000$ is the Sun's main sequence lifetime in Myr (Sackmann *et al.*, 1993). When a star ages to its main sequence lifetime, the stellar mass of the star is returned to the total available $M_{SF}$ at the corresponding cell, analogous to the formation of planetary nebulae.

*2.1.5 Star Formation Model*

We utilized the chemical evolution model of SFH developed by Snaith *et al.* (2014), who derive SFH from stellar abundances by fitting the abundance trends with age rather than the usual procedure of fixing the star formation prescription. This permits an unprecedented accuracy of the star formation history of the Galaxy in the first billion years as sampled in the solar vicinity. Their model obtained two distinct phases of the galactic inner disk formation (with a galactocentric radius < 8 kpc): thick and thin, and outer disk formation with similarly flat history, which posits that half of the stellar mass of the star is formed in the thick disk of Milky Way at approximately 4 Gyrs after its formation.

We prefer this approach because this model explicitly incorporated the significance of the galactic thick disk, a component that was recognized as early as 1986 (Gilmore and Wyse, 1986). While the thick disk represents the galactic stellar population in the most intense phase of star formation (Hopkins and Beacom, 2006; Madau and Dickinson, 2014), it is largely ignored in many star formation models when using the solar vicinity as a constraint that limited the intermediate metallicities of stars. For instance, the model of Fenner and Gibson (2003) and Naab and Ostriker (2006) lacked the consideration of explicit thick disk phase, resulting in the metallicity of the estimated stellar mass 10% less than the total stellar mass.

Using the Monte Carlo Realization technique, we allowed each star to form stochastically at each time step in respective disk regions (inner, outer), and computed the corresponding SFR from the SFH model (Snaith *et al.* 2014). At every time step (Myr), we allowed the stars to generate in a loop according to the corresponding SFR, with each star assigned to a random coordinate of (X, Y, Z). Then, our model obtained the calculated $M_{SF}$ (described in §*2.1.3*) from the corresponding cell at that location, and checks if there is sufficient available stellar mass for the star to form. If the criteria are met, the total $M_{SF}$ at that cell will be subtracted by the stellar mass of the newly formed star, resembling the conversion of gas into stars. This process of star formation will be repeated through the loop of the SFR from the SFH model, varying by different time phases of the Galaxy. By combining the theoretical approach of star formation prescription from the Schmidt-Kennicutt law with a SFH model derived from the accurate stellar observations, our model has greater power in that it respects the nature of star formation and does not rely too heavily upon a single model.

**2.2 Supernova**

Supernova explosion (SNe) rate is one of the major components of habitable models (Lineweaver *et al*., 2004; Gowanlock *et al*., 2011). A habitable environment is often described as a condition that is not frequently exposed to transient events, such as SNe. We simulated SNe, utilizing the galactic SN frequency estimated by Tammann *et al.* (1994). They modeled a SNe rate of $2.5 \times 10^{-2}$ yr$^{-1}$ and concluded that 85% of these SNes are from massive progenitors; see Tammann *et al.* (1994) for more detailed information. This is consistent with the theoretical estimates by Tutukov *et al.* (1992) who predicted the Type II supernova (SNII, explosion of a star with mass greater than 8 $M_\odot$) rate to be $1.96\text{-}3.35 \times 10^{-2}$ yr$^{-1}$. The frequency of Type Ia supernova (SNIa, explosion of a white dwarf in a binary system) events is estimated by Meng & Yang (2010), who find the value of 2.25 to $2.9 \times 10^{-3}$ yr$^{-1}$.

Gehrels *et al.* (2003) investigated the extinction effect of an averaged SNII and predicted that at a distance within 0.008 kpc, the flux of a SNII event is sufficient to deplete the ozone layer of any nearby planet and thus sterilize any land-based life in that planet. From this estimate, we assigned the sterilization distance for each SNII and SNIa according to their absolute magnitude distribution. With 46 observed SNII samples from an external galaxy, Miller and Branch (1990) address the distribution of SNII and conclude an average SNII magnitude in the B wavelength band ($M_B$, about 435nm wavelength) to be −16.89 and σ = 1.35. In addition, Richardson *et al.* (2016) developed the SNIa distribution from 239 SNIa samples with K-corrections and bias correction process and found the mean $M_B$ of −19.25 and σ = 0.5. From these distributions of SN, we propose a simple parameterization for the sterilization distance of each SNII and SNIa, which is inspired by the approach from Gowanlock *et al.* (2011):

$$D_{SNe} = d_{SNII} \times e^{-0.4(M_{SN} - M_{SNII})} \qquad (4)$$

where $D_{SNe}$ is the distance that will result in extinction of life from a given SN to a nearby life-bearing planet, and $M_{SN}$ is the absolute magnitude of the given SN. The constant $d_{SNII} = 0.008$ kpc denotes the extinction distance of an average SNII (Gehrels et al., 2003), and $M_{SNII} = -16.89$ is the mean absolute magnitude $M_B$ of SNII (Miller and Branch, 1990). We acknowledge that this parameterization may not obtain the most realistic, considering the complexity of sterilization mechanisms caused by SNe; however, it is reasonable enough to model SNe in our simulation.

When a sterilizing event occurs on a life-bearing planet, the ozone layer of that planet will be depleted and the land-based life within the planet will be removed. In our simulation, we allowed life to recur given a certain time period as $T_{min}$; the emergence of life will be defined as a Poisson process of abiogenesis, which will be presented in the next subsection.

## 2.3 Poisson Process of Abiogenesis

Abiogenesis, the creation of life from inanimate substances, is considered the most likely pathway towards the emergence of life (Wessen, 2010). Using mathematical modelling, we assume life is possible for Sun-like stars harboring Earth-like planets, on which life emerges with the Poisson process of abiogenesis. In this process, we assumed that separate events of abiogenesis do not influence each other, and that the probability of abiogenesis is simply a product of a uniform rate ($\lambda_A$, with unit $Myr^{-1}$) and time. In addition, we interpreted the complex and multi-path chemistry towards the accumulation of life as an ensemble, similar to previous work (e.g., Spiegel and Turner, 2012). Thus, the probability of life arising on a planet within time T is:

$$P_L = 1 - P_{Poisson}[\lambda_A, T, n = 0] = 1 - e^{-\lambda_A(T-T_{min})} \qquad (5)$$

where $P_{Poisson}[\lambda_A, T, n = 0]$ is the probability of obtaining zero successful events, $\lambda_A$ is the rate parameter, or the probability rate of abiogenesis in an Earth-like planet per unit time (Myr), and $T_{min}$ is the time period when a young Earth-like planet had a severe environment that precluded life by the time of its formation, uniformly chosen from a range of (0.1 to 1) Gyr. We prefer this approach of $T_{min}$ over a normal distribution or a specific chosen value because it will provide a relatively unbiased value and allow different Earth-like planets to vary, as the time period that strictly precludes life in Earth remains disputed. We recognize that this range itself is an assumption; however, as we currently lack a better understanding towards the origin of life, this range is reasonable enough to put a constraint on the results while allowing some variations to occur.

With this simplicity, we realize one major weakness: the parameter $\lambda_A$ may not be a uniform constant even when the planets are all Earth-like and possibly share common mechanisms for the development of life. The disparities of planetary environments over time may also influence our results. To address this problem, we include the parameter $T_{min}$, which responds to one of the major changes from a severe environment to a milder one. We argue that, as most of the stark changes in Earth came from the activities of life after the first success of abiogenesis event, those changes are irrelevant to this process of our model. Additionally, while Scharf and Cronin (2016) suggest that the rate of abiogenesis, $\lambda_A$, is heavily dependent on specific environmental conditions, we assume that Earth-like planets are likely to have a very similar rate of abiogenesis, $\lambda_A$.

Although previous work attempts to provide some estimations to $\lambda_A$ (e.g., Spiegel and Turner, 2012), we notice the value of this parameter varies greatly. To extend previous work of quantifying $\lambda_A$, we assume no particular value but simply present varying results with different values of $\lambda_A$ and evaluate their weights on a global scale of distributed stellar systems. We focus on the parameter of 1 and $10^{-6}$ Myr$^{-1}$, the upper and lower limits obtained from the range suggested by Spiegel and Turner (2012).

## 2.4 Sufficient time for the evolution of intelligence

We have insufficient evidence to ascribe a specific value for the timescale required for life to evolve into intelligence (denoted as $T_{evo}$, in Gyr). While Earth may provide a hypothetical inference for this parameter, we present the results from selected potential values. On Earth, $T_{evo}$ is approximately 3 to 3.5 Gyrs from the emergence of life to the current state where we start the SETI. Whether or not this is a typical timescale on other Earth-like planets is unknown, so we therefore selected three potential values: $T_{evo} \pm 0.2$ Gyr, where $T_{evo} = 1, 3, 5$.

We recognize the possible assumption made by Carter (2008) and Forgan (2009) that the process of evolving intelligence consists of sub-processes that each have a specific timescale. However, unless a particular evolutionary mechanism with sub-processes is suggested with sufficient evidence, it is not necessary to add complexity only to incorporate speculative distribution. Thus, we prefer to "integrate out" the possible detailed steps and treat the sub-processes as an ensemble of time period. Additionally, we account for the SNe explosion, and suggest the parameter $T_{evo}$ we described in this subsection corresponds to the time period that is free from transient events; when a nearly-sterilizing SNe threatens land-based life during the evolutionary process, we simply reset the process of abiogenesis and evolution on that planet, allowing them to recur as the planet ages.

## 2.5 Annihilation of Intelligence

While no evidence explicitly suggests that intelligent life will eventually annihilate themselves, we cannot *a priori* preclude the possibility of self-annihilation. As early as 1961, Hoerner (1961) suggests that the progress of science and technology will inevitably lead to complete destruction and biological degeneration, similar to the proposal by Sagan and Shklovskii (1966). This is further supported by many previous studies arguing that self-annihilation of humans is highly possible via various scenarios (e.g., Nick, 2002; Webb, 2011), including but not limited to war, climate change (Billings, 2018), and the development of biotechnology (Sotos, 2019).

To address these possibilities, we developed a simple probability parameter $P_{ann}$ to denote the probability of self-annihilation for galactic complex life. In each time step (1 Myr), we incorporate $P_{ann}$ to the development of complex life with Monte Carlo methods. We assume that this parameter remains uniformly constant over time, which lacks temporal variations or different social aspects from the civilizations and may result in a large variation. However, by no means do we attempt to estimate $P_{ann}$, as this parameter serves only to provide a qualitative understanding on possible outcomes of intelligent life, that we, as complex life, can potentially annihilate ourselves. We also utilize different values of $P_{ann}$ to present the varying results from extreme case scenarios (e.g., when $P_{ann}$ is set to 0 or 0.99) as means to examine its impact on the global prevalence of ETI in the Milky Way. We tested a range of values from 0 to 0.99, and selected the following for this parameter: 0, 0.5 and 0.99; the lower the value of $P_{ann}$, the greater the maximum number of potential galactic life. For all three values, we provided their spatial and temporal variations on the galactic prevalence of intelligent life. We excluded the value of 1 for the upper limit, simply due to the fact that we are still alive; a value of 1 would mean that the level of intelligent life in the entire Galaxy would be zero. On the other hand, a value of zero cannot be ruled out, as a civilization may become immortal (Grinspoon 2003, Nittler 2004).

## 3. Results and Discussion

This work presents a model of the Milky Way which simulates the evolution of galactic ETI and has produced a set of criteria to trace where and when complex life would occur.

### 3.1 Spatial-temporal analysis on the occurrence of ETI

**Table 1:** Each numerical value represents the maximum $Z_{ETI}$ of its spatial-temporal profile in Figure 1.

|  | $T_{evo}$ = 1 Gyr | | $T_{evo}$ = 3 Gyr | | $T_{evo}$ = 5 Gyr | |
| --- | --- | --- | --- | --- | --- | --- |
|  | $\lambda_A = 10^{-6}$ | $\lambda_A = 1$ | $\lambda_A = 10^{-6}$ | $\lambda_A = 1$ | $\lambda_A = 10^{-6}$ | $\lambda_A = 1$ |
| $P_{ann}$ = 0 | 7,811,780 | 8,729,415 | 4,566,340 | 5,362,915 | 2,832,970 | 3,598,260 |
| $P_{ann}$ = 0.5 | 7,805 | 10,455 | 2,159 | 2,880 | 1,117 | 1,510 |
| $P_{ann}$ = 0.99 | 80 | 105 | 30 | 40 | 25 | 25 |

We provided eighteen scenarios with galactic-scale resolution to examine potential outcomes dependent on varying parameters, plotted in Figure 1 and summarized in Table 1. For each key parameter, we investigated the quantity of ETI ($Z_{ETI}$) with spatial and temporal profiles. Figure 1 has the current time line (13 Gyrs) and our location at the current time (8 kpc, 13 Gyrs) indicated for reference

In Figure 1, a specific value of $\lambda_A$, 1 or $10^{-6}$ Myr$^{-1}$, was assigned to each vertical panel. For better comparison, we presented the results from different $\lambda_A$ with the same color-contour coding and found similar results. This is possibly due to the large number of Earth-like planets around the peak areas, and given enough time, life may be common there; such a significantly large sample of life resulted in little variation. Thus, we suggest that life is common elsewhere and the prevalence of ETI is not significantly dependent on the parameter $\lambda_A$. Within the time range developed by Spiegel and Turner (2012), the probability of life arising is likely not the factor behind the Fermi Paradox. Unless we speculatively assume a value for $\lambda_A$ far below the lower limit of the range developed by previous work, the number and prevalence of ETI is unlikely to alter significantly, which may discredit the Rare Earth hypothesis as a resolution to the Fermi Paradox.

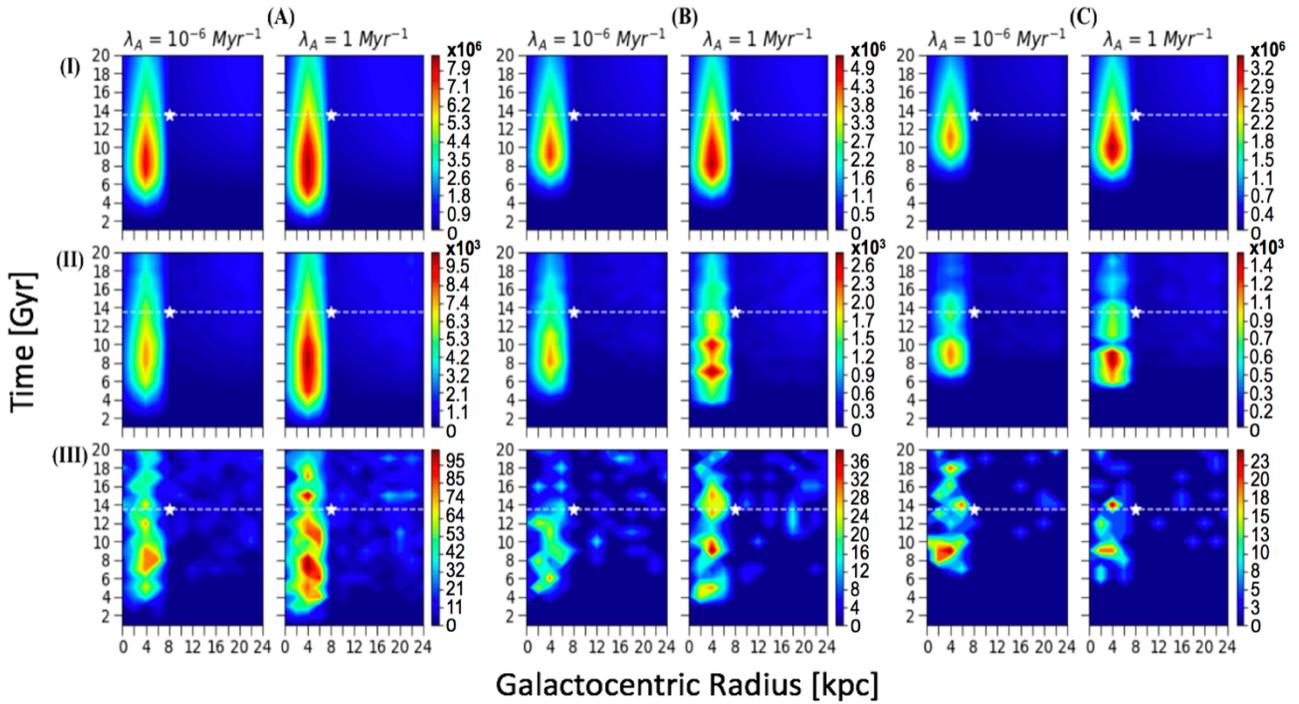

**Figure 1:** Spatial-temporal profiles for $Z_{ETI}$ (A) $T_{evo}$ = 1 Gyr, (B) $T_{evo}$ = 3 Gyrs, (C) $T_{evo}$ = 5 Gyrs; (I) $P_{ann}$ = 0, (II) $P_{ann}$ = 0.5, (III) $P_{ann}$ = 0.99, over 20 Gyrs. Each vertical individual panel corresponds to a specific value of $\lambda_A$, as shown in the title above. White star marks the current location of our sun at 8 kpc, and 13.5 Gyrs, and the white dash line represents our current time line of 13.5 Gyrs.

As discussed in our methods above, we used a range of $P_{ann}$ values from 0 to 0.99 for examining the impact of the potential of self-annihilation on the prevalence of ETI. The selected values of 0, 0.5, and 0.99 are plotted in horizontal panels (I), (II) and (III), respectively. As shown, the range of $Z_{ETI}$ varies significantly with different values of $P_{ann}$, suggesting that the quantity of galactic intelligent life is highly dependent on the likelihood of self-annihilation. One can interpret the significance of this parameter on the related hypothesis (Great Filter theory) that the probability of self-annihilation is possibly high, resulting in an extremely small fraction of ETI. We take no position in this argument; rather, our focus is to highlight the growth propensity for potential intelligent life in the Milky Way. We note that in panel (III), $P_{ann}$ = 0.99 shows that some proportions of the spatial-temporal variations deviate from general patterns due to its small size resulting from an extremely high possibility of self-annihilation; however, this will not invalidate our conclusions as: 1) the probability of 0.99 is an extreme case scenario and the actual probability of self-annihilation is unlikely to be so close to 1; and 2) the general peak location and how prevalence of ETI change spatially and temporally remain consistent in their spatial-temporal profiles.

Additionally, we consider the varying results with three potential $T_{evo}$. The vertical panels (A), (B), and (C) correspond to a timescale of 1, 3, 5 ($\pm$ 0.1) Gyrs, respectively, each plotted with the same scale for time and galactocentric radius. In panel (A), the peaks stretch a wider time period than other two panels because civilizations appeared earlier with a shorter timescale required for evolution. Results for growth propensity remain generally consistent, and the number range does not vary significantly compared to what is presented with different values for $P_{ann}$.

Though the range of $Z_{ETI}$ varies significantly with change in parameters like $P_{ann}$, peak location and spatial-temporal propensity for the prevalence of ETI remain precise throughout all scenarios. From Figure 1, we can develop a comprehensive picture of where and when potential complex life has formed: at the inner Galaxy between 2-8 kpc, with the center focus of the peak 4 kpc from the Galactic Center and time around 8 Gyrs, and number decreasing monotonically outward. The $Z_{ETI}$ immediately dropped at the boundary of the galactic inner disk (8 kpc), and remained a low number density throughout the galactic outer disk. We noted that our location is not within the region where most ETI occur, as our sun is located outside of the boundary of the inner Galaxy (see the white star in Figure 1); our location is merely too far from potential complex life. We suggest for the SETI to be further into the inner Galaxy, ideally towards the annulus 4 kpc from the galactic center.

In Figure 1, the quantity of intelligent life also decreases with time after the peak, possibly due to an outburst of ETI at the peak time, gradually decreasing as Sun-like stars age to their main sequence lifetime, and may eventually reach an equilibrium between birth and death of intelligent life. In our model, this explanation is tested to be true in Figure 2 by plotting the spatial-temporal change of normalized $Z_{ETI}$ over a stretch of 50 Gyrs. The parameter $P_{ann}$ is set to 0 in this plot, suggesting the only factors resulted in this equilibrium are star formation and death, and sterilizing SNe. Namely, this latter equilibrium state also remarks the peak point as an unusual outburst of ETI over time. In particular, Figure 2 marks the equilibrium state at around 20 Gyrs, and the prevalence of ETI remained identical throughout the next 30 Gyrs.

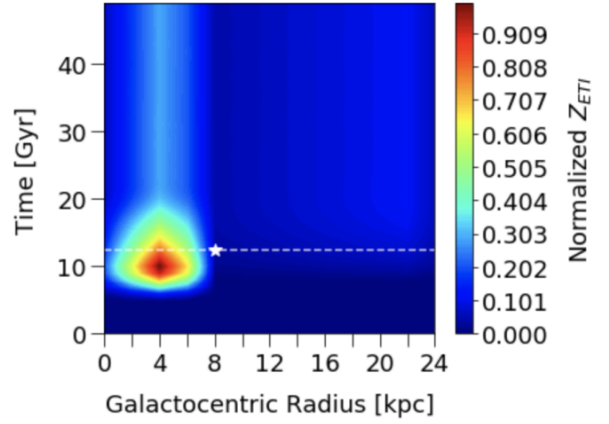

**Figure 2**: Normalized spatial-temporal profile for $Z_{ETI}$ over 50 Gyrs, when $\lambda_A = 10^{-6}$ Gyr$^{-1}$, $P_{ann} = 0$ and $T_{evo} = 5$ Gyrs.

Further, to assess the impact of sterilizing SNe on the $Z_{ETI}$, we presented the sterilized number of ETI with similar spatial-temporal analysis, plotted in Figure 3. We presented the results when $T_{evo}$ is set to 3 Gyrs, simply because one collection of plots is enough to see the correlation between the sterilizing SNe frequency and the prevalence of ETI. However, we presented results from other varying parameters as a means to compare the plots with the corresponding spatial-temporal profiles from vertical panel (B) of Figure 3. We found the quantity of ETI increases with the frequency of sterilizing SNe. When the $Z_{ETI}$ reaches the maximum, the SNe event that sterilized complex life also reaches a maximum, and we can conclude that the significantly higher number density of habitable stars outweighs the impact of more frequent sterilizing SNe events.

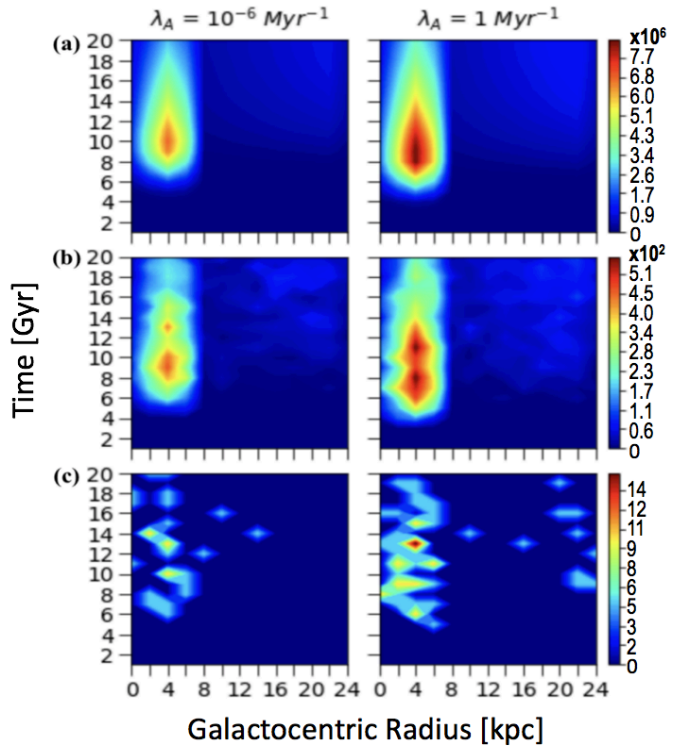

**Figure 3**: The sterilizing SNe spatial-temporal profiles (a) $P_{ann} = 0$, (b) $P_{ann} = 0.5$, (c) $P_{ann} = 0.99$. Color bar represents the number of sterilized ETI from SNe.

## 3.2 The effect of intelligence annihilation parameter on age distributions

To examine how old the ETI are at the current time of the Milky Way, we plotted age distributions for varying values of the key parameters ($T_{evo}$, $\lambda_A$, $P_{ann}$), and found that $P_{ann}$ observably influences the results of age distribution. Therefore, we ran the model with $P_{ann}$ values ranging from 0 to 0.99, and selected three resulting plots in Figure 4. We selected the value 0, 0.001, and 0.1 because each changes the age distribution greatly. When $P_{ann}$ is set to 0 in panel (a), the ages of ETI vary significantly from 0 - 10 Gyrs. However, this variation immediately drops when $P_{ann}$ is set to 0.001 in panel (b), with ages only varying from 0 to 2.5 Gyrs. This suggests that $P_{ann}$ must be extremely low to permit age variations to occur. More, when increasing $P_{ann}$ to 0.1 in panel (c), all complex life become less than 0.06 Gyr in age, and the majority stay younger than 0.01 Gyr. Since we cannot preclude the high possibility of annihilation, Figure 4 suggests that most of potential complex life within the Galaxy may still be very young.

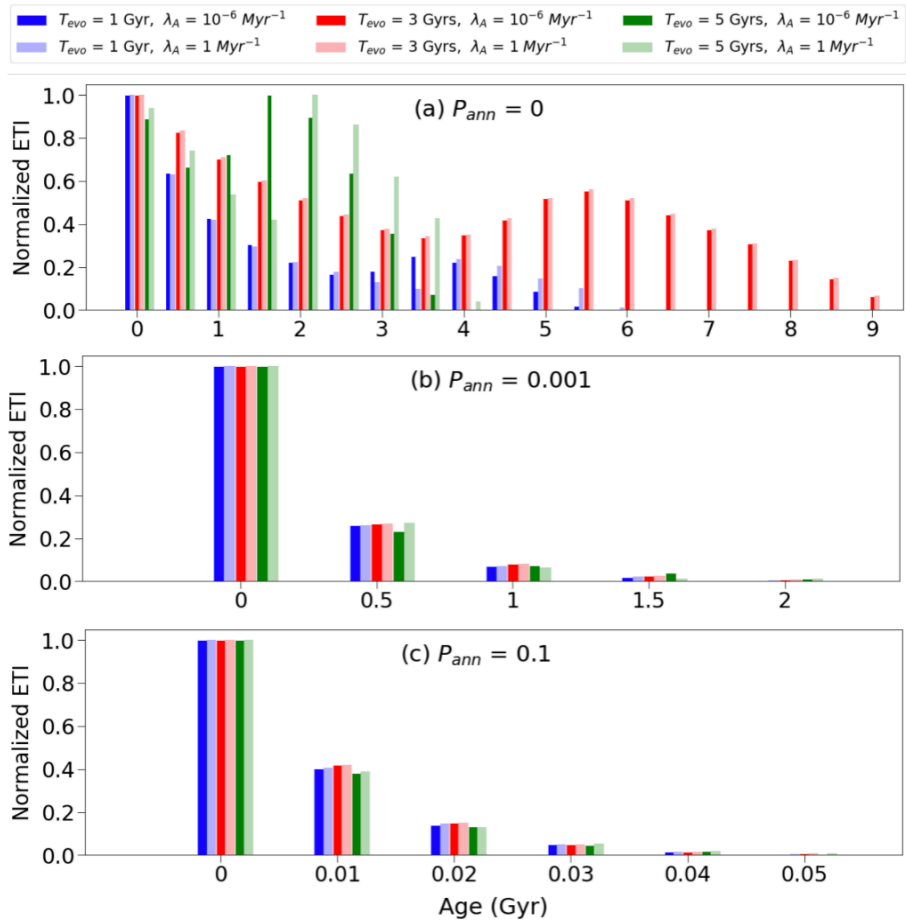

**Figure 5:** Age distributions (a) $P_{ann}$ = 0, (b) $P_{ann}$ = 0.001, (c) $P_{ann}$ = 0.1. The range of each bin for panels (a) and (b) are 0.5 Gyr, and the range of bin for panel (c) is 0.01 Gyr. The numbers in x-axis ticks represent the starting value of its bin range. All panels share the same legend of color coding on the top.

## 3.2 Discussion

The precise peak location of the inner Galactic disk we suggested is consistent with many previous studies (Forgan and Rice, 2010; Gowanlock et al., 2011; Morrison and Gowanlock, 2015; Forgan *et al.*, 2017). Our work produced a probabilistic peak location regarding the galactocentric radius and time, with galactic-scale resolution stretching over 20 Gyrs. This unprecedented stretch of time and space allows us to compile a comprehensive picture of when and where potential ETI evolve, and presents all potential profiles of galactic complex life, and their impacts on the level of ETI with varying values of key parameters. Our model covers the entire Galaxy, permitting us to discard the constraint of their limited galactic radius range from the work by Morrison and Gowanlock (2015); a constraint prevented their model from obtaining a precise peak location, and their result lay beyond the lower limit of the range (< 2.5 kpc).

We note that it is difficult to put a further constraint on the number of ETI due to the inherent uncertainty of any specific parameter values; however, our main conclusions are not altered. When more future analysis on the emergence and development of ETI are applied, we can develop an improved range for the quantity of ETI with spatial and temporal variance. We also note that our simulation for gas and conversion of gas to stars are coarse, with a static Milky Way spatial hash table lacking the considerations of stellar kinematics; however, we only focus on the growth propensity of ETI in a galactic scale, which are not heavily associated to the migration of stars.

## 4. Conclusions

A galactic model with resolution of individual Earth-like planets has been developed to analyze where and when potential intelligent life is most abundant throughout time, and how old the majority of complex life are at the current time in the Milky Way. A precise analysis for the propensity of potential ETI is formed, with a set of criteria including: the formation of two distinct phases of the Galaxy, recent observational analysis on the prevalence of Earth-like planets, sterilizing SNe that threaten land-based life, Poisson process of abiogenesis, different evolutionary timescales and potential self-annihilation of life. We utilized three parameters, $\lambda_A$, $T_{evo}$, $P_{ann}$, to trace the origin, evolution and development of life in the Milky Way, and assess the spatial and temporal variations for the quantity of ETI.

Extending the time to 20 Gyrs, regardless of change in key parameters, resulted in the peak for the number of ETI across all galactic regions around 4 kpc, and time at 8 Gyrs. The number of ETI decreased spatially and temporally from the peak. Even if we vary the evolutionary timescale to be 2 Gyrs less or greater than what was required on Earth, and with different values of $\lambda_A$ and extreme case scenarios of $P_{ann}$ (0 and 0.99), our conclusion of the precise location and time for the peak is not altered. Our results suggest that the quantity of intelligent life does not always increase with time; in fact, our model predicts that after the peak occurs, the number of ETI starts to decrease monotonically with time, and this propensity remains throughout the next 6.5 Gyrs. More, our results show that the level of ETI will eventually reach an equilibrium between birth and death of intelligent life at approximately 20 Gyrs. Further investigation for the number of sterilizing SNe events suggests that the prevalence of ETI reaches its maximum despite having the highest frequency of sterilizing SNe events by the location of the peak.

We also concluded that at the current time of the study, most intelligent life in the Galaxy is younger than 0.5 Gyr, with values of probability parameter for self-annihilation between 0 - 0.01; with a relatively higher value of the annihilation parameter ($\geq 0.1$), most intelligent life is younger than 0.01 Gyr. As we cannot assume a low probability of annihilation, it is possible that intelligent life elsewhere in the Galaxy is still too young to be observed by us. Therefore, our findings can imply that intelligent life may be common in the Galaxy but is still young, supporting the optimistic aspect for the practice of SETI. Our results also suggest that our location on Earth is not within the region where most intelligent life is settled, and SETI practices need to be closer to the inner Galaxy, preferably at the annulus 4 kpc from the Galactic Center.

Additionally, we found the potential self-annihilation to be highly influential in the quantity of galactic intelligent life, suggesting another possible answer to the Fermi Paradox; if intelligent life is likely to destroy themselves, it is not surprising that there is little or no intelligent life elsewhere. Further, the probability of abiogenesis was not found to influence the quantity of intelligent life within the range developed by previous works, which supports the optimistic perspective of life being common in the Galaxy.

The results of our model yield a range of possible quantities for intelligent life over time. We assert that the intelligent life will always be most abundant approximately 4 kpc from the galactic center, peaking at time around 8 Gyrs, decreasing monotonically from that peak point, and that a majority of potential intelligent life is still young. The exact number of the intelligent life estimated here is not the focus of our work; rather, it is instead the development of a statistical, comprehensive galactic picture tracing the potential growth propensity of intelligent life over a course of ~20 billion years.

**Acknowledgments**

This research was supported by the Jet Propulsion Laboratory, California Institute of Technology, under the contract with NASA. We acknowledge the partial funding support from the NASA Exoplanet Research Program NNH18ZDA001N.**Data Statement**

The data underlying this article are available in the article and in its online supplementary material. For additional questions regarding the data sharing, please contact the corresponding author at <Jonathan.H.Jiang@jpl.nasa.gov>.

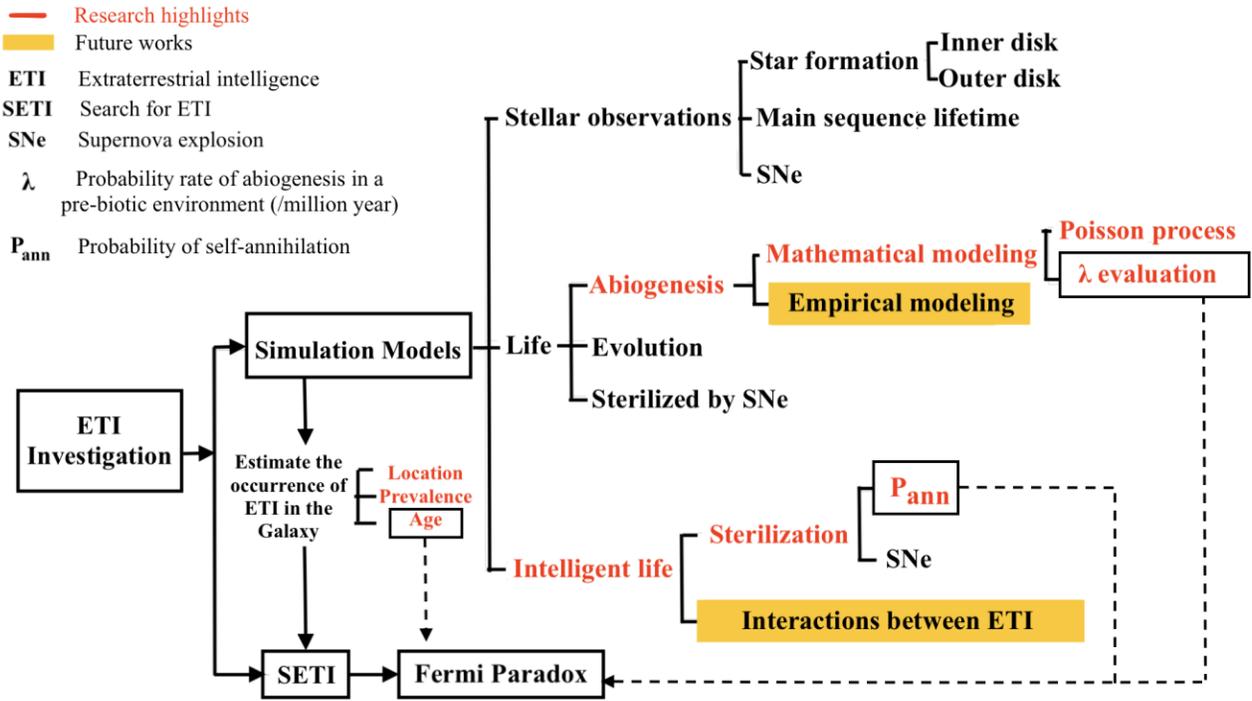

**Appendix Figure 1:** An illustrated diagram of approach of key components for ETI investigation as well as our research highlights and potential future works.